\documentclass[%
reprint,
superscriptaddress,
 amsmath,amssymb,
 aps,
 pra,
]{revtex4-1}

\usepackage{graphicx}
\usepackage{dcolumn}
\usepackage{bm}
\usepackage{hyperref}
\usepackage[mathlines]{lineno}

\newcommand{\pbar}{$\bar{\text{p}}$\,\,} 
\newcommand{\Hbar}{$\bar{\text{H}}$\,\,} 
\newcommand{\Hbarplus}{$\bar{\text{H}}^+$\,\,}

\begin{document}


\title{Optical trapping of anti-hydrogen towards an atomic anti-clock}

\author{P.~Crivelli}
\email[]{crivelli@phys.ethz.ch}
\affiliation{Institute for Particle Physics and Astrophysics, ETH Zurich, 8093 Zurich, Switzerland}
\author{N.~Kolachevsky}
\affiliation{P.N. Lebedev Physical Institute, Leninsky prospekt 53, Moscow, Russia}

\date{\today}
\begin{abstract}

The Anti-Matter Factory at CERN is gearing up, commissioning of the Extra Low ENergy Antiprotons (ELENA) ring is ongoing and the first anti-protons are foreseen to circulate in the decelerator very soon. The unprecedented flux of low energy antiprotons delivered by ELENA will open a new era for precision tests with antimatter including laser and microwave spectroscopy and tests of its gravitational behaviour. 
Here we propose a scheme to load the ultra cold anti-hydrogen atoms that will be produced by the GBAR experiment in an optical lattice tuned at the magic wavelength of the 1S-2S transition in order to measure this interval at a level comparable or even better than its matter counter part. 
This will provide a very accurate test of Lorentz/CPT violating effects which can be parametrised in the framework of the Standard Model Extension.  

\end{abstract}

\pacs{Valid PACS appear here}
\maketitle


\section{\label{sec:level1}Introduction}

The recent discovery of the Higgs boson at LHC was another great triumph of the Standard Model (SM) of particle physics.  This model is providing incredibly accurate predictions validated by the experiments. However, despite its success, the SM cannot account for the origin of dark matter and for the matter-antimatter asymmetry observed in our Universe. In fact if the SM would be the correct description of Nature we would not even exist since matter and anti-matter would have annihilated during the Universe expansion right after the Big Bang leaving behind a desolated universe populated only by radiation. 
To produce such an asymmetry the famous Sakharov conditions should be invoked.  CP violation in the leptonic sector might be the key and different experiments are planning to test this possibility in the near future.  Additional scenarios have been put forward such as CPT violation as e.g. in the context of the Standard Model Extension (SME)  \cite{Kostelecky}. A very vibrant experimental activity in this direction is ongoing with the goal to test the very foundation of the SM and quantum field theory which rests on CPT invariance. The projected sensitivity of the CPT tests of the aforementioned measurements can be parametrised in terms of the absolute precision which can be used for comparison of different systems. Furthermore, since CPT tests are also tests of Lorentz symmetry within conventional field theories, those might also shed light on the development of a unified theory of gravity and quantum mechanics \cite{Kostelecky}.

 Among those experiments, anti-hydrogen is a blossoming field that sprout with the measurement by the TRAP collaboration of the antiproton g-factor \cite{Gabrielse:1990tc,Gabrielse:1995tu} and the first observation of $\bar{\text{H}}$ at the CERN Low Energy Anti-protons Ring (LEAR) \cite{LEARHBAR}. Those motivated the construction of the Anti-proton Decelerator (AD) facility which allowed the production of antihydrogen ($\bar{\text{H}}$) at low energies \cite{ATHENA2002, ATRAP2002}. The formation of $\bar{\text{H}}$ was achieved by mixing trapped positrons and antiprotons plasmas in a nested Penning--Malmberg trap \cite{NestedTrap}. By refining this technique, $\bar{\text{H}}$ can now be  trapped magnetically for more than 1000 s \cite{ALPHA2010,ALPHA2011, ATRAP2012}. This led already to interesting results such as a test of \Hbar neutrality and a first test of gravity on anti-matter. Furthermore, this important milestone led to the first observation of the 1S-2S transition of anti- hydrogen \cite{ALPHA2017} and a detailed measurement of the transition line shape will provide one of the best test of CPT invariance by comparison with normal hydrogen. The hyperfine splitting of \Hbar has also been measured very recently at a level of four parts in $10^{-4}$\cite{ALPHA2017HFS}.
New improved measurements of the charge to mass ratio and the magnetic moment of the antiproton have been performed using Penning traps \cite{BASE} and of the antiproton- electron mass ratio determined with spectroscopic studies of antiprotonic helium \cite{HoriScience2016}.

Steady progress towards a very precise hyperfine splitting measurement  of anti-hydrogen \cite{ASACUSA2014HFS} is also being made.  A method to form a $\bar{\text{H}}$ beam was recently demonstrated making use of a CUSP trap \cite{ASACUSA2010}. 

Moreover, anti-hydrogen is being used to test the gravitational behaviour of antimatter. A first direct limit has been inferred on the gravitational acceleration of antimatter by releasing the $\bar{\text{H}}$ atoms from the magnetic trap \cite{ALPHAGBAR}. Improvements on this setup, comprising laser cooling or a vertical magnetic trap \cite{ALPHAGBARPlus}, could lead to a test of the effect of gravity on antimatter with a precision of 1\% or better. With the same goal two proposals have been approved at CERN \cite{Aegis, GBar}. Both experiments are progressing and they are planning to form anti-hydrogen via charge exchange of positronium (Ps) with anti-protons:\\
\begin{eqnarray}\label{eqn:HBARProd}
& \text{Ps}+\bar{\text{p}}\to\bar{\text{H}}+\text{e}^-.
\end{eqnarray}
Production of normal hydrogen via charge exchange of protons with positronium has been demonstrated by M. Charlton et al. \cite{Charlton1997}.
The same mechanism has already been proven to produce $\overline{\text{H}}$ in Rydberg states ($\overline{\text{H}}^*$) by the ATRAP collaboration using a two step charge exchange process, i.e. formation of Rydberg positronium with positrons impinging on Cs$^*$ atoms and subsequent formation of $\overline{\text{H}}^*$ \cite{ATRAP2004,ATRAPJMol1_2016}. 
The cross sections for the charge exchange reactions were calculated by different authors \cite{Mitroy1995}-\cite{Charlton2016} and good agreement has been found with the available experimental data. 

\section{Measurement of 1S-2S transition in hydrogen and anti-hydrogen} 

The most precise determination of the 1S-2S transition in hydrogen has been done by the group of T. W. H\"ansch at the Max Planck Institute of Quantum optics at the impressive level of a relative fraction uncertainty of $4.2\times10^{-15}$ \cite{MPQ}. This was achieved using a 6 K cryogenic beam interacting with a laser standing wave at 243 nm to perform Doppler free two photon spectroscopy. A laser radiation of 6-13 mW was injected in a build up Fabry-Perot resonator. The circulating  intra-cavity power varied between 150 and 300 mW to characterize the AC Stark shift.  The atoms excited in the 2S state would cross an electric field in which they would quench to the 2P state via Stark mixing emitting a Lyman alpha photon. The number of these photons detected as a function of the laser frequency allows one to build the resonance curve and thus extract the 1S-2S energy interval.  The hydrogen beam was chopped at 160 Hz by a mechanical wheel allowing to select atoms in the low energy tail of the Maxwell Boltzmann distribution by means of time-of flight. Atomic velocity groups with the central velocities from 70 m/s to 110 m/s (depending on the time bin) were used for the analysis in order to reduce the second order Doppler shift and the time-of-flight broadening at the price of reducing the statistics. 
The transition frequency was extrapolated to zero 243 nm power to compensate for the AC Stark shift which for the given intra-cavity power is at a level of a few hundred Hz.
The measured line width was at a level of 2 kHz and the transition frequency was determined with an uncertainty of 10 Hz. 

Even though \Hbar beams are now available  \cite{ASACUSA2014HFS}, it would seem far beyond any imagination for today's technology to get a cold beam of $10^{16}$ anti-hydrogen atoms per second as done for hydrogen.  Therefore the ATRAP and ALPHA collaborations opted for magnetic trapping of \Hbar atoms to perform precise 1S-2S spectroscopy on this transition as it was pioneered at MIT with normal hydrogen \cite{Claudio1996}. 
Very recently the ALPHA collaboration reported the first observation of the 1S-2S transition in anti-hydrogen. Their result is compatible with CPT invariance at a relative precision of $2\times10^{-10}$ \cite{ALPHA2017}. For the trap configuration in the ALPHA experiment of 1T depth the line broadening is at the level of 30-40 kHz, therefore, a measurement of the line shape should allow to improve their current precision by at least one order of magnitude. 
Laser cooling as realised in Amsterdam \cite{Amsterdam} for magnetically trapped H with a pulsed Lyman alpha laser will reduce the anti-atoms temperature to 10-20 mK thus the trapping field could be lowered by more than an order of magnitude resulting in a line width at the kHz level. Therefore the ultimate relative uncertainty of this method should be at a level of $10^{-12}-10^{-13}$. 
To further improve this measurement in order to be comparable with its matter counter part ultra cold \Hbar atoms should be produced.

\section{Production of  ultra cold \Hbar atoms (10 $\mu$K) }\label{sec:MagTrap}
The GBAR experiment is planning to form ultra cold anti-hydrogen (10 $\mu$K) using a two step charge exchange process. The anti-hydrogen produced using the charge exchange reaction from the anti-protons coming from the ELENA decelerator \cite{ELENA} with Ps (see Eq.\ref{eqn:HBARProd}) will interact with the same Ps target forming anti-hydrogen ions through the charge exchange reaction:
\begin{eqnarray}
& \text{Ps}+\bar{\text{H}}\to\bar{\text{H}}^++\text{e}^-.
\end{eqnarray}
The cross sections of this process were computed by different authors \cite{Comini2013} but have not been measured yet.

The idea to exploit anti-ions in order to produce ultra cold anti-hydrogen was first proposed by J. Walz and  T. H\"ansch \cite{Walz2004}.
The advantage of \Hbarplus ions is that those can be sympathetically cooled with normal matter ions since the Coulomb repulsion prevents their annihilation.

After being formed the anti atoms will then be guided electro-statically to a Paul trap in which a crystal of laser cooled Be$^+$ ions has been prepared. Via Coulomb interaction with the Be$^+$, the \Hbarplus will cool down in less than 1 ms. An anti-ion will then be transferred to a precision trap to form a pair with a single Be$^+$ ion. The ion-anti-ion pair will then be cooled via Raman transitions down to 10 $\mu$K \cite{HilicoWAG2013}. At this point a laser pulse detaches the excess positron producing ultra cold neutral anti-hydrogen. The photo detachment  will be done close to the threshold (the binding energy is 0.76 eV) in order not to impart a significant momentum to the anti-hydrogen atom. In GBAR, the laser pulse defines the time $t_0$ for the start of the free fall measurement. The stop is given by the detection of the charged pions originating by the annihilation of the anti-proton in the anti-hydrogen atom on the walls of the free fall chamber thereby with the free fall time the effect of the gravity of antimatter can be determined and $\bar{g}$ extracted to a level better than 1\%. A more detailed description of the GBAR experiment and its current status can be found in \cite{GBar}.

After the photo-detachment the low field seekers ultra cold anti-hydrogen atoms (50\% of the initial population) could be captured in a magnetic trap by overlapping this to the precision ion trap before the photo detachment of the ions occurs. 
The trap depth for a given magnetic field strength $B$ is $U= 0.6 B\, {\rm{K/T}} =60 B\, \mu\rm{K/G}$ thus few Gauss are sufficient to confine the neutral \Hbar atoms after photo-detachment. The magnitude of these fields is the typical one used to manipulate the Zeeman splitting of the ground state sub levels and  will thus not perturb the trapping of the ion pairs.
If e.g. a Ioffe trap \cite{pritchard} would be used, spectroscopy of \Hbar could already provide an improved result since the atoms will be a factor 100 times colder than what one might be able to achieve by laser cooling.  This reduces the TOF broadening by a factor 10, the second order Doppler broadening by a factor 30 and the Zeeman broadening by two orders of magnitude. The line width will be at a level of 100 Hz, therefore, even with few atoms one could expect to reach an uncertainty better than $10^{-13}$.

\section{Optical trapping of \Hbar at the magic wavelength}
The magic wavelength for hydrogen, at which the lowest-order AC Stark shifts of the 1S and 2S states are equal, has been recently calculated to be 514.6 nm \cite{Kawasaki2015, Jentschura}. The value of the magic angular frequency  $\omega_\text{M}$ is a function of several transition frequencies and matrix elements that need to cancel the Stark shift of the ground and 2S states and therefore depends on the Rydberg constant  $\hbar \omega_\text{M}\approx0.177 R_\infty$ \cite{Jentschura}.
From the recent observation of the 1S-2S \Hbar transition \cite{ALPHA2017} one can estimate that the magic angular frequency for \Hbar compared to H might differ by not more than 60 kHz. This could be detected if the magic wavelength for H (to note that this has not been yet measured) is also used for \Hbar. 

An optical trap can be formed by tightly focusing the far detuned laser.
The depth of the trap can be calculated using \cite{Grimm}:
\begin{equation}
U_{dip}=-\frac{3\pi c^2}{2 \omega_0^3}\Big(\frac{\Gamma}{\omega_0-\omega}+\frac{\Gamma}{\omega_0+\omega}\Big)I,
\end{equation}\label{eq:Udip}
where $\omega_0=2\pi c/\lambda$ is the 1S-2P transition frequency which is at the Lyman-alpha $\lambda=121.5$ nm, $\omega=2\pi c/\lambda$ the frequency of the far detuned laser at the 514.6 nm magic wavelength. The frequency line width of the 1S-2P transition is $\Gamma=1/\tau_{2P}$ where $\tau_{2P}=1.6$ ns is the lifetime of the 2P state in hydrogen.
The intensity for a Gaussian beam is given by:
$I=2 P/(\pi w_0^2)$
where $w_0$ is the laser beam waist. 
For a standing wave lattice, the trap depth obtained with Eq.\ref{eq:Udip} has to multiplied by a factor of four. 
The scattering rate of the atoms with the trapping laser is given by \cite{Grimm}:
\begin{equation}
\Gamma_{sc}=-\frac{1}{\hbar}\Big(\frac{\omega}{\omega_0}\Big)^3\Big(\frac{\Gamma}{\omega_0-\omega}+\frac{\Gamma}{\omega_0+\omega}\Big)U_{dip}.
\end{equation}\label{eq:Gammasc}

A commercial solution for a laser outputting 5 W at the magic wavelength was discussed and could be produced by Toptica \cite{toptica}. A fiber amplifier will be seeded with an ECDL at 1029.2 nm that is than frequency doubled. About 3 W of laser light at 514.6 nm will be injected in a high finesse (nearly) confocal resonator to generate around 3 kW circulating power. The concave mirrors with a radius of curvature $R$=50 mm produce a beam waist of 10 $\mu$m so that the intensity in the trap centre would be around 2 GW/cm$^2$. 
Such an intensity will provide a trap depth of 26 mK which is more than 300 times the recoil energy of $E_{\rm rec}=72$ $\mu$K and thus will result in robust trapping. 
The intensity on the mirrors would be around 0.3 MW/cm$^2$. This value is less than the one obtained in the 486 nm enhancement cavity used for positronium spectroscopy \cite{hype15}. In this experiment, up to 0.5 kW power is routinely build up with an intensity on the mirrors of 0.4 MW/cm$^2$ without observing any degradation over days of running. Only above 0.7 MW/cm$^2$ the mirrors are damaged. 

The AC Stark shift of the hydrogen ground state at the magic wavelength is 221.6 Hz/(kW/cm$^2$), and the slope of the AC Stark shift at the magic wavelength under a change of the driving laser frequency is 0.2157 Hz/[GHz (kW/cm$^2$)] \cite{Jentschura}. Therefore to keep the line width of the transition below 10 Hz, the laser should be stabilised to about 1 kHz which does not present any challenge for an ECDL. 
All the other sources of line broadening will be at a the same level as described in detail in the next Section, therefore, after the magic wavelength will be found experimentally  (this has to be determined to 1 kHz) by measuring the 1S-2S transition frequency as a function of the magic wavelength to minimize the AC Stark shift, few atoms will be enough to determine this transition at a level of $10^{-15}$ (corresponding to an uncertainty of 10 Hz).

\section{Loading of the optical trap}

One possibility to load the ultra cold anti-atoms in the standing wave lattice is to first capture them in a magnetic trap as mentioned in Sec.\ref{sec:MagTrap}. However, to achieve a tight localisation of the atoms a steep gradient around the trap minimum is required before turning on the trapping laser. A simple anti-Helmholtz trap seems the best choice for this purpose. To localise the atoms down to 10 microns the field gradient should be 100 G/mm which seems feasible since neutral atoms magnetic trapping with gradients up to $3 \times 10^4$ G/mm has already been demonstrated \cite{Vuletic}. The drawback is that Majorana losses due to non-adiabatic transitions at zero of the field in the trap center will limit the lifetime of the \Hbar atoms in the trap \cite{bergeman} thus a fast on/off switching of the magnetic field in less than 1 ms time scale will be required. This could be achievable using a low inductance superconducting magnet energised discharging a large capacitor through an IGBT \cite{CanJ2009} but the implementation of such a scheme requiring cryogenics is very challenging.

A more promising path to load the optical trap is to use a pulsed laser for the photo-detachment at threshold of the anti-ions instead than a CW one as planned originally in the GBAR experiment \cite{GBARProposal}. A pulsed laser would allow for the prompt capture of the neutral anti-atoms in the time scale of the trapping laser power build up. This would not be possible in the CW regime since turning on the optical trap will immediately photo-detach the anti-ions and give a large recoil velocity of about 500 m/s to the \Hbar atoms. 
The probability for the photo-detachment ($P_{PD}$) can be estimated with:
\begin{equation}
 P_{PD}= \frac{\sigma P}{\pi \omega_0 E_\gamma} \Delta t, 
\end{equation} 
where $\sigma=3.8\times10^{-16} \Delta E^{3/2}\,\rm{cm}^2$ is the photo-detachment cross section above the energy threshold $\Delta E$ (in eV) \cite{lykke}. The energy of the photons $E_\gamma \simeq$ 0.76 eV corresponds to the anti-ion binding energy and $P$, $\omega_0$ and $\Delta t$ are the laser power, beam waist and pulse duration respectively. 
For an energy above threshold of 15 $\mu$eV as planned in GBAR,  a photo-detachment probability close to 100\% can be achieved with a laser pulse of 30 mJ and 5 ns focused down to 15 $\mu$m. 
Commercial laser systems \cite{quantel} based on OPO can deliver the required power at 1640 nm with a laser line width of 0.01 cm$^{-1}$ \cite{LaserSpec} corresponding to 1.2 $\mu$eV. 

The ponderomotive force that the ions experience due to the inhomogeneous oscillating electromagnetic field from the laser is given by:
\begin{equation}
F_{q}= \frac{q^2}{4m \omega^2} \nabla |{E_0}|^2 , 
\end{equation} 
where $q$ and $m$ are the ion charge and mass respectively and $\omega$ and $E_0$ are the field frequency and amplitude.
The spatial intensity distribution for a linearly polarized Gaussian laser beam in cylindrical coordinates in the focal plane (the field gradient is maximum in the radial direction) reads:
\begin{equation}
I(r,0)=  |{E_0}|^2  = I_0exp\Big(\frac{-2r^2}{w^2_0}\Big)
\end{equation} 
where $w_0$ is the laser beam waist.
For a laser pulse length $\tau$ with a time dependence of the form $exp(-t^2/\tau^2)$, the acceleration is thus given by:
\begin{equation}
\ddot r(t)=\frac{q^2}{m^2\omega^2}\frac{I_0r(t)}{w_0^2}exp\Big(\frac{-2r(t)^2}{w^2_0}\Big)exp(-\frac{t^2}{\tau^2}).
\end{equation}
Assuming that the atoms do not move significantly during the laser pulse, one can set $r(t)=r$ and integrate the equation above to obtain an analytical solution \cite{Eichmann}. The maximal force experienced by the atoms is at $r=w_0/2$ and therefore the maximal velocity is:
\begin{equation}
v_{\text{max}}=\frac{q^2 I_0\sqrt{\pi}Erf(1)exp(-0.5)}{2m^2 \omega^2 w_0}\tau
\end{equation}
thus for the given parameters of the photo-detachment laser this will be of the order of $v_{\text{max}}=0.02$ m/s and can therefore be neglected. 

When the photo-detachment laser is fired, the trap laser will be switch on and the power build up in the resonator will start. 
This will require about 1 $\mu$s. 
Since after the photo-detachment the anti-ions will move with a velocity of the order of 1 m/s during this time those will have traveled not more than a 1 $\mu$m which is much smaller than the trapping laser waist.  To note is that the anti-ions are initially localised inside the ion trap to better than 1 $\mu$m.
To be able to start the trap laser build up simultaneously with the photo-detachment the cavity should be pre-locked. Using the 514.6 nm or the fundamental, even assuming that few tens of nW could be locked to the resonator, would result in a high probability of photo detachment. Therefore, one should use an additional laser at 2058.4 nm. This wavelength is above the photo-detachment threshold and the second harmonic can be phase locked to the fundamental frequency of the trap laser. 
Since the cavity line-width for the parameters of the resonator is of the order of 250 kHz a commercial DFB laser system with few mW seems a good choice for this purpose. 

\begin{figure}[h!]
\centering
\includegraphics[width=0.4\textwidth]{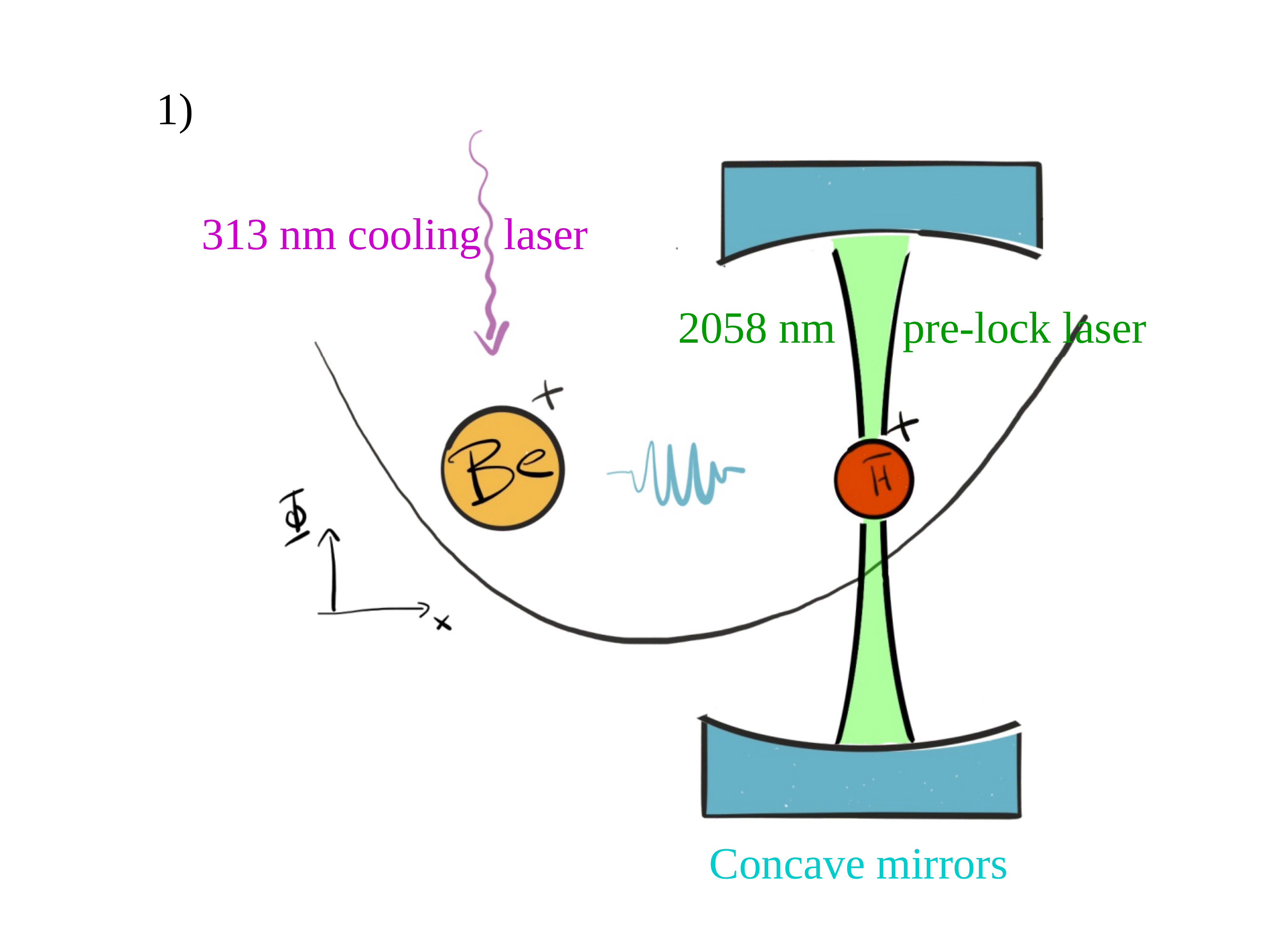}
\includegraphics[width=0.4\textwidth]{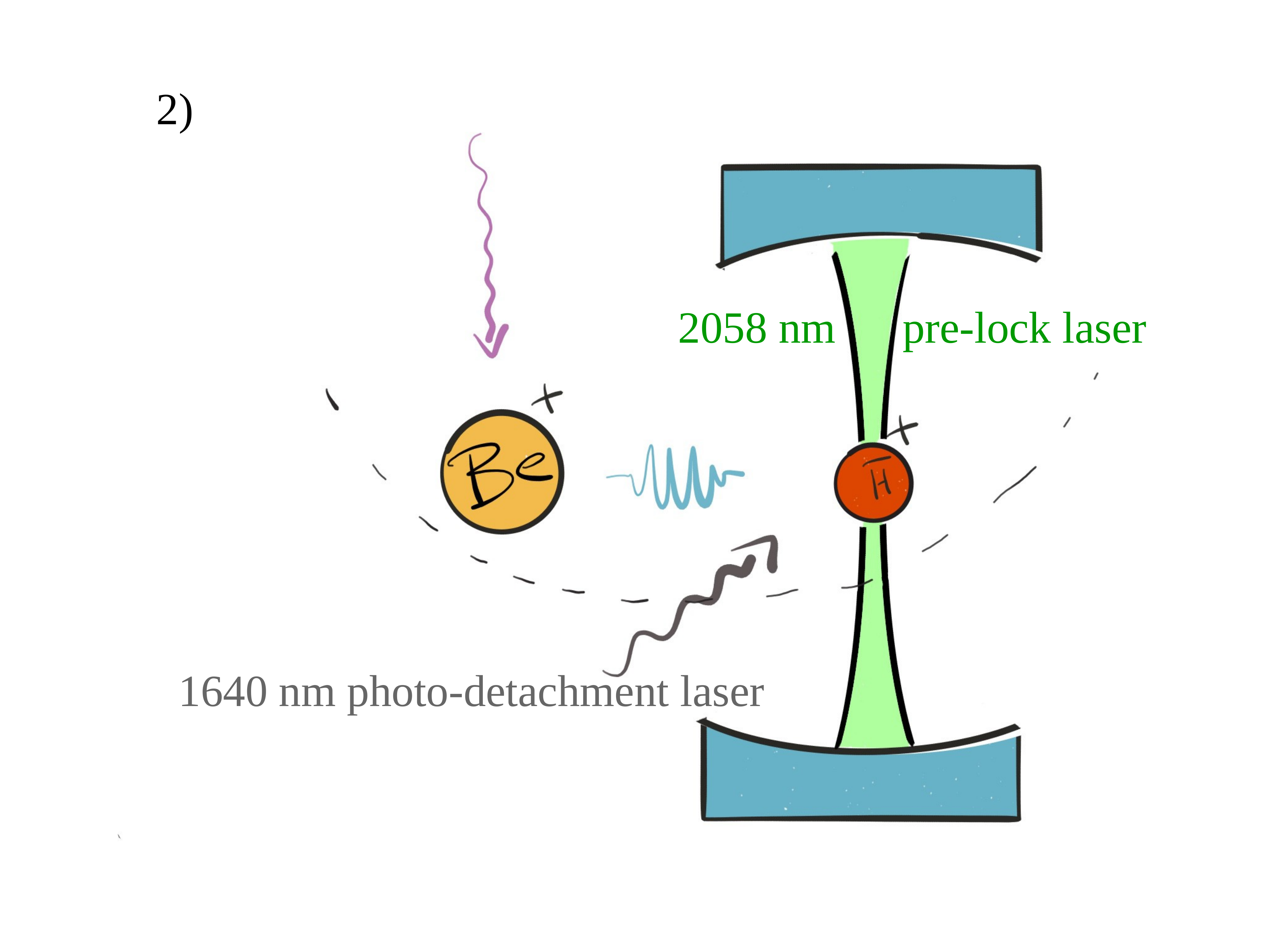}
\includegraphics[width=0.4\textwidth]{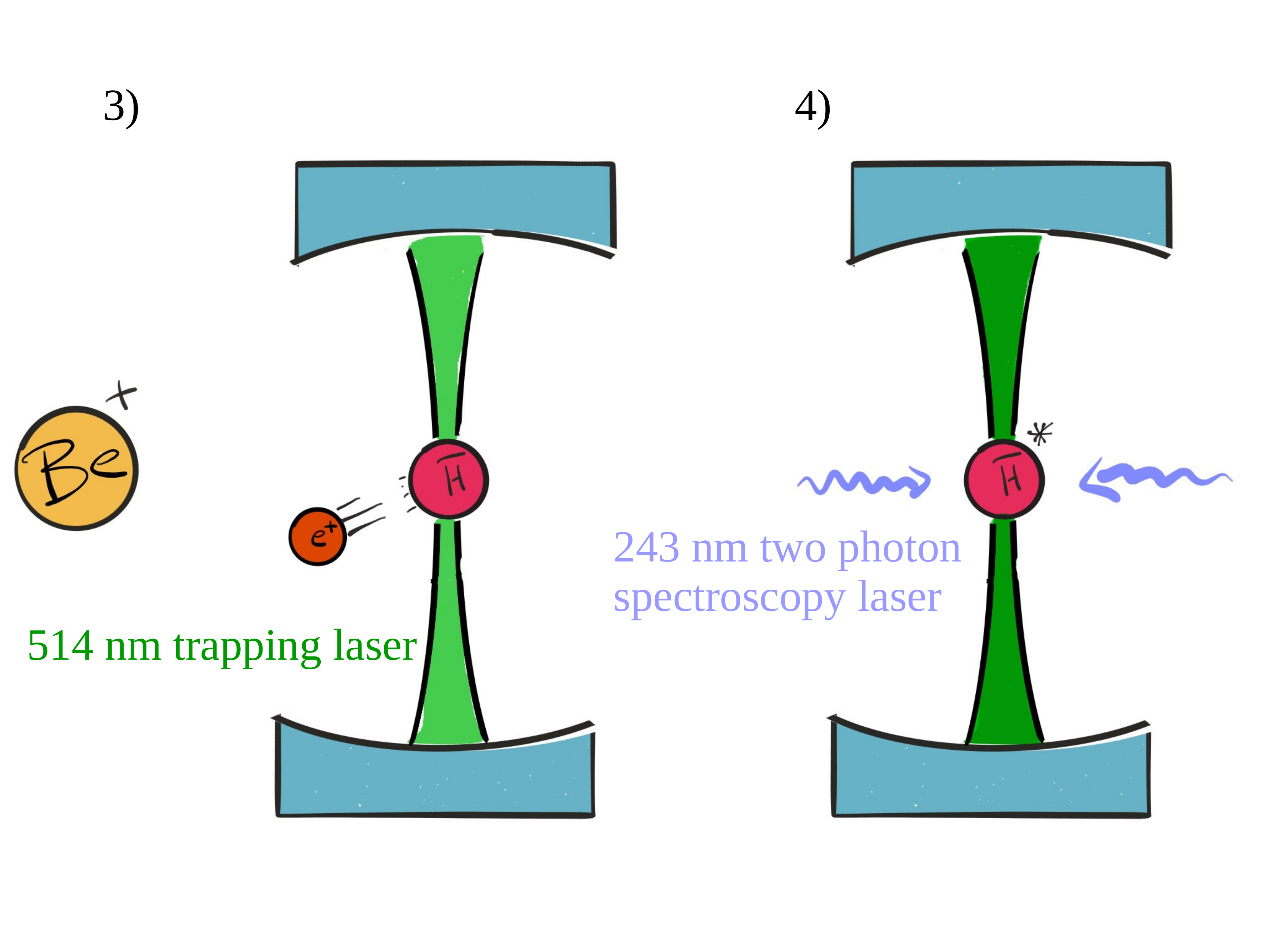}
\includegraphics[width=0.4\textwidth]{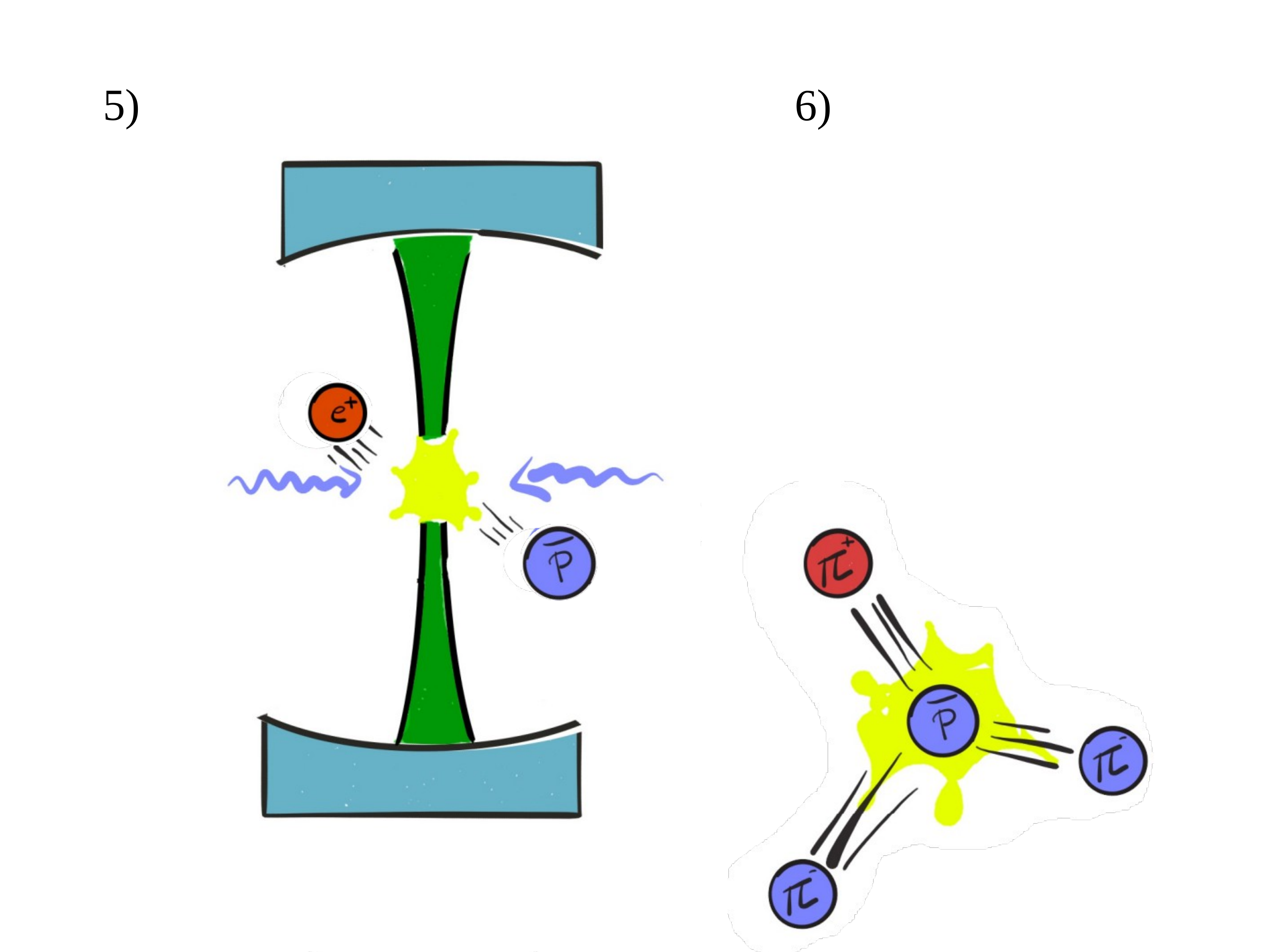}
\caption{Sketch of the loading sequence and spectroscopy.}
\label{Loading}
\end{figure}

Once the atoms are trapped the 243 nm laser will be used to excite the \Hbar in the 2S state. The signal will be detection of the anti-protons annihilations released from the trap after photo-ionization of the anti-hydrogen atoms in the same spectroscopy laser as shown schematically in Fig. \ref{Loading}.  
To summarise the trap loading sequence and spectroscopy is:
\begin{enumerate}
\item The Be$^+$- \Hbarplus pair is trapped and cooled and the 2058 nm laser pre-locked to the resonator.
\item The photo-detachment laser pulse is fired.
\item The 514 nm laser for the optical trap is turned on to build up the power in the Fabry Perot and the ion trap fields are switched off, thus, the Be ion will be removed and no electric fields will be present.
\item The two photon spectroscopy 243 nm laser is turned on. 
\item The \Hbar atom is photo-ionized. 
\item The pions from the anti-protons annihilations on the vacuum chamber are detected. 
\end{enumerate} 

\section{Expected rate and precision on the \Hbar 1S-2S transition in the optical trap}

The expected excitation and photo-ionisation rates are found by solving the optical Bloch equations for the two-photon transition \cite{Haas2006}:
\begin{subequations}
\begin{eqnarray}
\frac{\partial}{\partial t}\rho'_{gg}(t)&=&-\Omega(t)\,
{\rm{Im}}(\rho'_{ge}(t)) + \gamma_{\rm 2S}\rho'_{ee}(t),\\\nonumber \rule{0pt}{5ex}
\frac{\partial}{\partial
t}\rho'_{ge}(t)&=&-{\rm{i}}\,\Delta\omega(t){\rho}'_{ge}(t)+
{\rm{i}}\frac{\Omega(t)}{2}({\rho}'_{gg}(t)-{\rho}'_{ee}(t))\\
&&-\frac{\gamma_{\rm i}(t)+\gamma_{\rm 2S}}{2}{\rho}'_{ge}(t),\\
\rule{0pt}{4ex}
\frac{\partial}{\partial t}\rho'_{ee}(t)&=&\Omega(t)
{\rm{Im}}(\rho'_{ge}(t))-(\gamma_{\rm i}(t)+\gamma_{\rm 2S})\rho'_{ee}(t),
\end{eqnarray}
\label{EOMdensityMatrix}
\end{subequations}
where $\rho_{gg}$, $\rho_{ee}$ are the components of the density matrix for the ground and excited states while $\rho_{eg}$ is their superposition. 
The term $\gamma_{2S}=1/\tau_{2S}$ takes into account the spontaneous emission of the 2S state back into the ground state with the lifetime $\tau_{2S}=122$ ms.
The Rabi frequency is given by:
\begin{equation}
\Omega(t)=2(2\pi\beta_{ge})(1+m_e/M_p)^3I(t).
\end{equation}
Here $m_e$ and $M_p$ are the electron and the proton masses respectively.  $I(t)$ is the laser intensity seen by the atoms as a function of time.
For the 1S-2S transition in hydrogen the coefficient is $\beta_{ge}=3.68111\times10^{-5}$~Hz(W/m$^2$)$^{-1}$ \cite{Haas2006}. The ionization rate is:
\begin{equation}
\gamma_{\rm i}(t)=2\pi\beta_{\rm ioni}(e)(1+m_e/M_p)^3 I(t)
\end{equation}
with the ionisation coefficient $\beta_{\rm ioni}(e)=1.20208\times10^{-4}$ Hz\,(W/m$^2$)$^{-1}$\cite{Haas2006}.
The excitation detuning $\Delta\omega(t)$ can be expressed as:
\begin{equation}
\Delta\omega(t)=2 \omega_L-2 \pi \nu_{eg}- 2\pi (\Delta \nu_{ac}(e)-\Delta \nu_{ac}(g))+\Delta \omega_{2DS},
\end{equation}
where $ \omega_L$ is the laser detuning, $\nu_{eg}$ the transition frequency, the frequency shift due to the AC stark shift induced by the exciting laser field defined as:
\begin{equation}
\Delta \nu_{ac}(e)= \beta_{ac}(e) (1+m_e/M_p)^3 I(t)
\end{equation}
for the ground state and likewise for the excited one $\Delta \nu_{ac}(g)$.
Where the coefficients are $\beta_{ac}(e)=-2.67827\times 10^{-5}$ Hz/(w/m$^2$)$^{-1}$ and $\beta_{ac}(g)=1.39927\times10^{-4}$ Hz/(w/m$^2$)$^{-1}$\cite{Haas2006}.
The second order Doppler shift is:
\begin{equation}
\Delta \omega_{2DS}= - 2 \pi \nu_{eg}\frac{1}{2}\frac{v^2}{c^2},
\end{equation}\label{eq:DS2}
where $v$ is the atom velocity and $c$ the speed of light. 

To describe the full dynamics of the system in greater detail, other processes which are not considered in the given optical Bloch equations, such as non resonant excitations,  514 nm two photons ionization, heating due to lattice intensity fluctuations, should be included. However, those are not expected to have a significant impact on the experiment.

\begin{figure}[h!]
\centering
\includegraphics[width=0.45\textwidth]{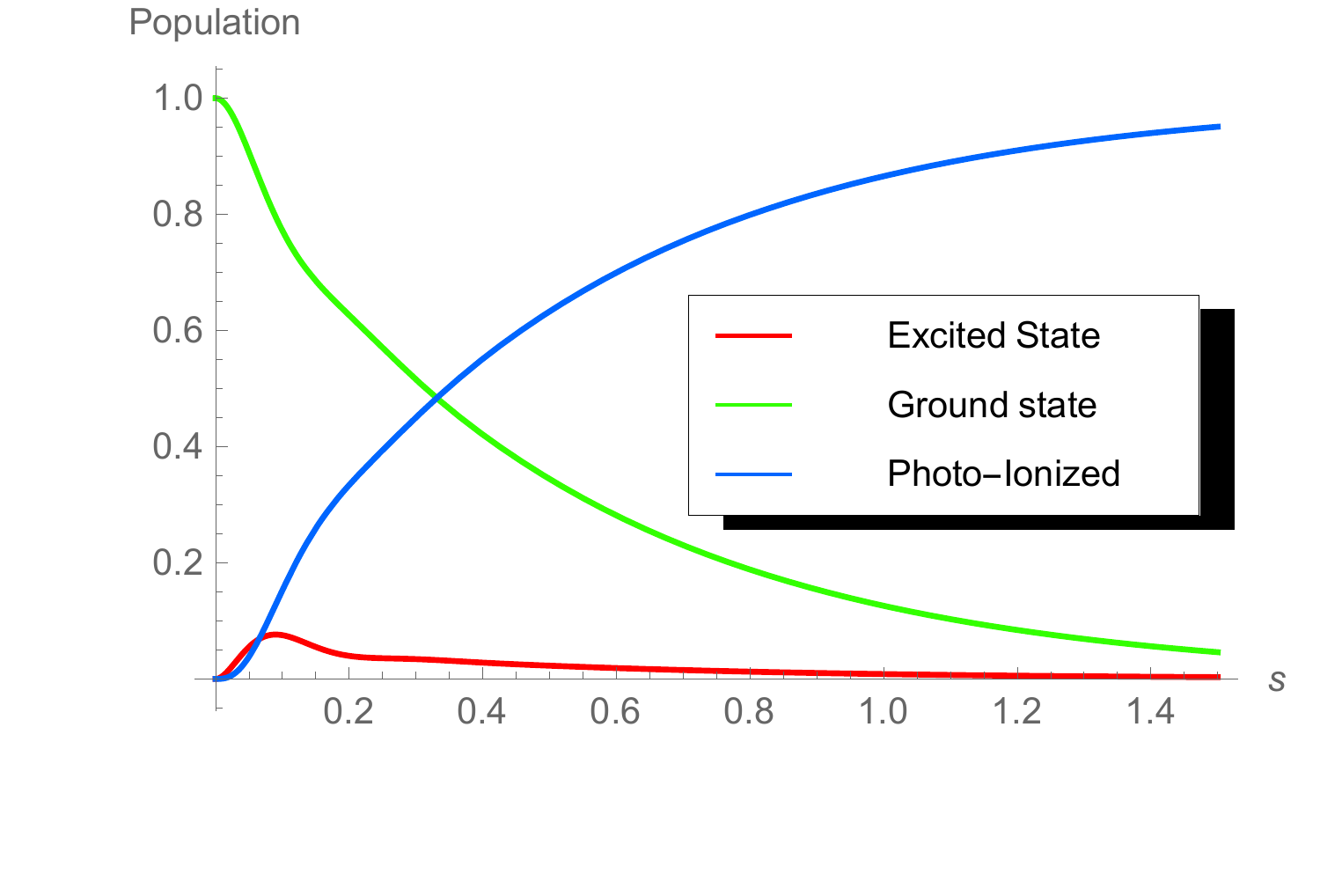}
\caption{Evolution of the population in the ground state (red curve), excited state (green curve) and the photo-ionized atoms (blue curve) as a function of the exposure time to the 243 nm laser light at the constant intensity of $1.6\times 10^{-4}$ W/cm$^2$.}
\label{Population}
\end{figure}

The results of the solution of the Bloch equations are shown in Fig. \ref{Population}. The beam waist of the laser ($\omega_0$=200 $\mu$m) was chosen to be larger than the trap region  in order to keep the AC stark broadening at the sub-Hz level. The AC Stark shift for the 1 mW circulating power will be around 3.6 Hz and by monitoring the circulating power can be corrected below the Hz level.   
 For the given power and beam waist the probability for photo-ionisation will be close to unity in a 1.5 s time window (see Fig. \ref{Population}). Once the \Hbar is photo-ionised, the \pbar and the positron will leave the trap.
 The atoms in the lattice will heat via scattering with the 514.6 nm trap laser at a rate that can be estimate with Eq.\ref{eq:Gammasc} to be of the order of $\Gamma_{sc}=4$ s$^{-1}$. The total heating results in an increase of the atom energy of $2E_{\rm rec}$ in a time $\Gamma_{sc}^{-1}$. After 1.5 s and for the given parameters, the velocity of the atoms will increase to about $v=6$ m/s and therefore the second order Doppler shift can be estimated with Eq.\ref{eq:DS2} to be at a level of 0.5 Hz. 
The signal is the detection of the charged pions from the anti-protons annihilations on the walls of the vacuum chamber.  As demonstrated by the ALPHA experiment this detection technique is quite efficient and the background from cosmic rays can be suppressed at a very high level \cite{ALPHA2017}.  They reported a 37\% detection efficiency for 300 s exposure time with a signal to noise ratio of 4. For a 1.5 s time window, the GBAR tracker will detect anti-protons annihilations with an efficiency of 54 \% and a signal to noise ratio of 250 (this can be improved at the expenses of the signal efficiency) \cite{DiPhD}.  Even higher efficiency and signal to noise ratios might be achieved with a more sophisticated analysis based on deep learning \cite{Sadowski:2017ilo}.

The GBAR experiment aims to produce 1 trapped \Hbarplus per ELENA cycle (100 s). Therefore the expected signal rate per day (we assume 8 hours duty cycle) can be estimated with:
\begin{equation}
R=\epsilon_d \cdot  N_{\bar{\text{H}}^+} \simeq 144  
\end{equation}
where $\epsilon_d=0.5$ is the detection efficiency. 
Although this experiment is extremely challenging, the estimated rate shows that one should be able to acquire enough statistics to split the line to the required level in one day allowing, thus, for careful studies of systematic effects. 	

An additional systematic to be considered is the Zeeman shift due to small relativistic corrections of the magnetic moment of the bound electron which depends on the bound state energy. In the presence of a magnetic field this result in a shift of the 1S-2S transition frequency which can be estimated with: 
\begin{equation}
\nu_Z= \frac{\alpha^2 \mu_B}{4 h} B=18 B \,{\rm Hz/G}
\end{equation}
where $\alpha$ is the fine structure constant, $\mu_B$ the Bohr magneton and $h$ the Planck constant. 
Therefore, the Zeeman shift due the Earth magnetic field will be at a level of 10 Hz and should be corrected for.
Although the ion trap fields will be switched off, some stray fields should be carefully taken into account and those should be suppressed to the level of 10 mV/cm in order not to introduce a DC Stark shift.
A summary of all the estimated contributions for about 150 detected atoms on resonance to the uncertainty is given in Table \ref{tab:systematics}. The main contribution is expected from the determination of the magic wavelength which should be known to about 1 kHz. The expected line with is about 10 Hz (FWHM) to which the main contribution is the stability in frequency of the trap laser.

\begin{table}[b!]
\begin{center}
\begin{tabular}{ccc}
\hline
\hline
 & $\sigma$ [Hz] & $\sigma$/f$_{1S-2S}$ [10$^{-15}$] \\
\hline
Statistics &$< $1 &  $<$ 0.4    \\
Zeeman shift & $< $1&  $<$ 0.4    \\
2nd order Doppler shift &0.5  & 0.2  \\
AC Stark shift & $<$ 1 &   $<$ 0.4  \\
DC Stark shift & 1.5 &   0.6 \\
Magic wavelength &$<$10 & $<$4   \\
\hline
Total & 10.3 & 4.2   \\
\hline
\hline
\end{tabular}
\end{center}
\caption{Estimated uncertainty budget for the anti-hydrogen 1S-2S frequency measurement for $f_{1S-2S} \simeq 2.466 \times 10^{15}$ Hz.}
\label{tab:systematics}

\end{table}

\section{Conclusions}
A scheme to measure the anti-hydrogen 1S-2S transition at a level that would be comparable and even super-seed its matter counter part is proposed.   Such a measurement might be feasible in the near future in the context of the GBAR experiment at CERN thanks to the ELENA ring, the recent upgrade of the AD, and the installation of an intense positron source based on a 10 MeV LINAC.  This experiment is the first step towards the development of an optical anti-clock which would provide a formidable tool to test Lorentz/CPT violating effects and anti-matter properties in general.

\begin{acknowledgments}

P. C. gratefully acknowledge M. Doser and E. Widmann for the very stimulating discussions and A. Antognini, J. P. Brantut, C. L. Cesar, D. Comparat, T. Donner, L. Hilico, F. Nez, E. Peik, R. Pohl and B. Radics for their very valuable comments and suggestions and L. Gerchow for his help with the manuscript. 
The work of P. C is supported by the Swiss National Science Foundation under the grant numbers 162794/166286 and by ETH Zurich. The work of N.K. is supported by the joint DFG-RFBR grants (HA 1457/12-1 and 17-52-12016).  

\end{acknowledgments}

\end{document}